\documentstyle[pre,aps,floats,epsf]{revtex}
\epsfclipon
\begin{document}
\draft
\wideabs{

\title{Off-lattice noise reduction and the ultimate scaling of DLA in two
    dimensions}
\author{Robin C. Ball$^1$, Neill E. Bowler$^1$, Leonard M. Sander$^2$, Ell\'ak
    Somfai$^{3,1}$}
\address{$^1$Department of Physics, University of Warwick, Coventry, CV4 7AL,
    England}
\address{$^2$Michigan Center for Theoretical Physics, Department of Physics, 
    University of Michigan, Ann Arbor, Michigan, 48109-1120}
\address{$^3$Instituut-Lorentz, Universiteit Leiden, Niels Bohrweg 2, 2333 CA
    Leiden, Netherlands.}
\maketitle

\begin{abstract}
Off-lattice DLA clusters grown with different levels of noise reduction are
found to be consistent with a simple fractal fixed point.  Cluster shapes and
their ensemble variation exhibit a dominant slowest correction to scaling, and
this also accounts for the apparent ``multiscaling'' in the DLA mass
distribution.  We interpret the correction to scaling in terms of renormalized
noise.  The limiting value of this variable is strikingly small and is
dominated by fluctuations in cluster shape.  Earlier claims of anomalous
scaling in DLA were misled by the slow approach to this small fixed point
value. 
\end{abstract}

\pacs{PACS numbers: 64.60.Ak, 61.43.Hv}

} 

\section{Introduction}

Since its introduction in 1981, the Diffusion-Limited Aggregation model of
Witten and Sander\cite{witten} has been a paradigm of self-organised scaling
behaviour in irreversible growth. However, even after twenty years, there is
still controversy about its scaling properties; many authors have claimed, for
example, that DLA clusters do not scale as simple fractals, but instead have
various anomalous features.  In this paper we give data on DLA clusters with
noise reduction which enables us to refute conclusively the basis of these
claims of anomalous scaling. We will show that the apparent anomalies
arise from a slowly decaying correction to scaling which can be associated
with the level of intrinsic growth fluctuations, as suggested in \cite{somfai}.
The analysis of these corrections to scaling gives us
considerable insight into the asymptotic behaviour of DLA, i.e. the DLA fixed
point.  

In (off-lattice) DLA a cluster is rigid and stationary, growing from one seed
particle by accretion at first contact of $N$ mobile diffusing hard sphere
particles.  The diffusing particles are sufficiently dilute so that they can
be taken to arrive one at a time.  We consider the distribution of where
growth (by
deposition) occurs at a given cluster size.  The average radius of deposition
is defined by $R_{\mathrm dep}= \langle r \rangle$, where $r$
is the the distance of deposition from the center of the cluster.  There is no
controversy that $R_{\mathrm dep} \propto N^{1/D}$, consistent with a simple
fractal of dimension $D = 1.71$ for large clusters in two dimensions.  However
the spread of the deposition radius is thought to show anomalies.  Plischke
and R\'acz\cite{plischke} introduced the penetration depth, $\xi$, the
standard deviation of radius of deposition of a given cluster, and claimed
that it scaled differently from $R_{\mathrm dep}$.  More recently, Davidovitch
\emph{et al.} \cite{davidovitch} considered the standard deviation of the
cluster average radius \emph{across the ensemble} of clusters, $\delta
R_{\mathrm eff}$, and claimed  that it was asymptotically negligible compared
to the mean. Another anomalous feature that has been claimed of DLA is
multiscaling \cite{coniglio,amitrano}: the fractal dimension of the cluster is
said to depend on the distance (relative to the cluster radius) from the
center. We will examine these claims using finite size scaling with the help
of noise reduction and show that none of them hold. We find that DLA is
consistent with simple scaling, and the apparently anomalous scaling can all
be explained by a slow correction to scaling.

\section{Off-lattice noise reduction}

Noise reduction for the {\em lattice version} of DLA has been introduced
\cite{lattice-noisered,barker} with the aim of suppressing the shot noise of the
individual incoming particles. When growing at lower noise levels, the
clusters achieve more asymptotic behavior at smaller sizes: a prime example
of this is that the lattice effects show up earlier. These lattice effects on
noise reduced clusters (or without noise reduction on very large clusters) are
quite strong, so in order to avoid them, any analysis of large scale DLA
clusters has to be made off-lattice.

\begin{figure}
\epsfxsize=1\columnwidth
\epsfysize=0.414\columnwidth
\centerline{\epsffile{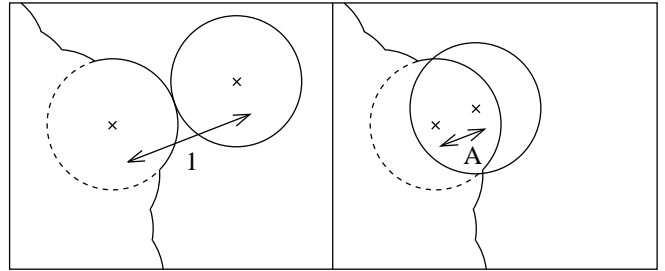}}
\medskip
\caption{The noise-reduced DLA algorithm.  A particle is allowed to diffuse
freely until it contacts a cluster particle.  The diffusing particle is moved
onto the cluster particle, reducing the distance between their centers by a
factor $A$.}
\label{how dla works}
\end{figure}

\begin{figure}
\epsfxsize=1\columnwidth
\epsffile{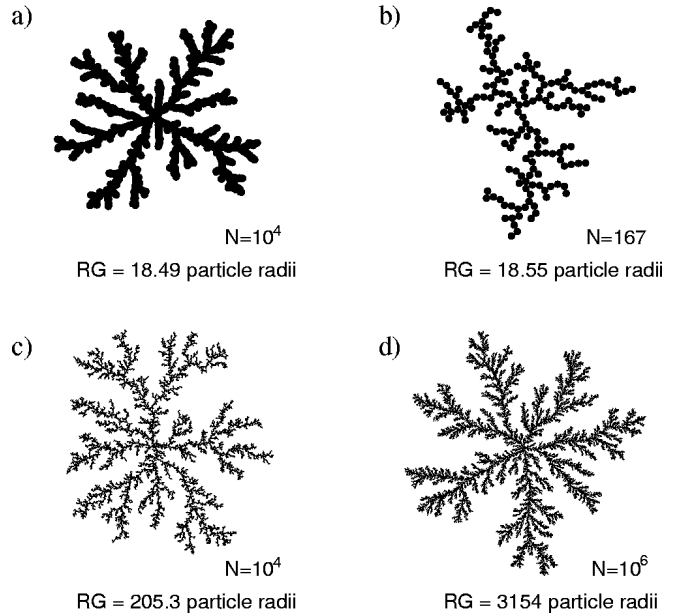}
\smallskip
\caption{Cluster a) grown with noise reduction $A=0.03$ is compared with
clusters grown without noise reduction, $A=1$: b) has same gyration radius, c)
has same number of particles and d) has same correction to scaling properties
(e.g.  relative penetration depth). The numbers shown are the particle number
and gyration radius.}
\label{Clusters}
\end{figure}

In our version of noise-reduced off-lattice DLA the particles diffuse freely
until they contact a particle in the cluster, just as in the original model.
However, on contact with a particle of the existing cluster, the diffusing
particle is moved into that particle by a factor of $A$, at which point it is
irreversibly stuck (see Fig.~\ref{how dla works}).  This means that shallow
bumps are added to the cluster, and that we must add $1/A$ particles on top of
one another to protrude the growth by a particle diameter.  A cluster grown
with this method of noise reduction is shown in Fig.~\ref{Clusters}. Another
way to do noise-reduction of this type was introduced by Stepanov and
Levitov\cite{stepanov}, who generalized the method of iterated conformal maps
\cite{hastings,davidovitch} to add shallow bumps.

\section{Finite size scaling}

Growing clusters at a variety of levels of noise reduction gives us a very
clear picture of the finite size scaling effects in DLA.  We grew 1000 DLA
clusters to 1,000,000 particles with noise reduction levels of $A=0.3, 0.1,
0.03, 0.01$, and 4000 clusters with $A=1$ as well as 25 clusters with
$A=0.001$.  At various points in the growth, 100,000 probe particles were
fired at each cluster to measure its properties.  In the following
measurements the center of the cluster was taken naturally as the center of
mass of these probe particles (``center of charge''). 

\begin{figure}
\epsfxsize=1\columnwidth
\centerline{\epsffile{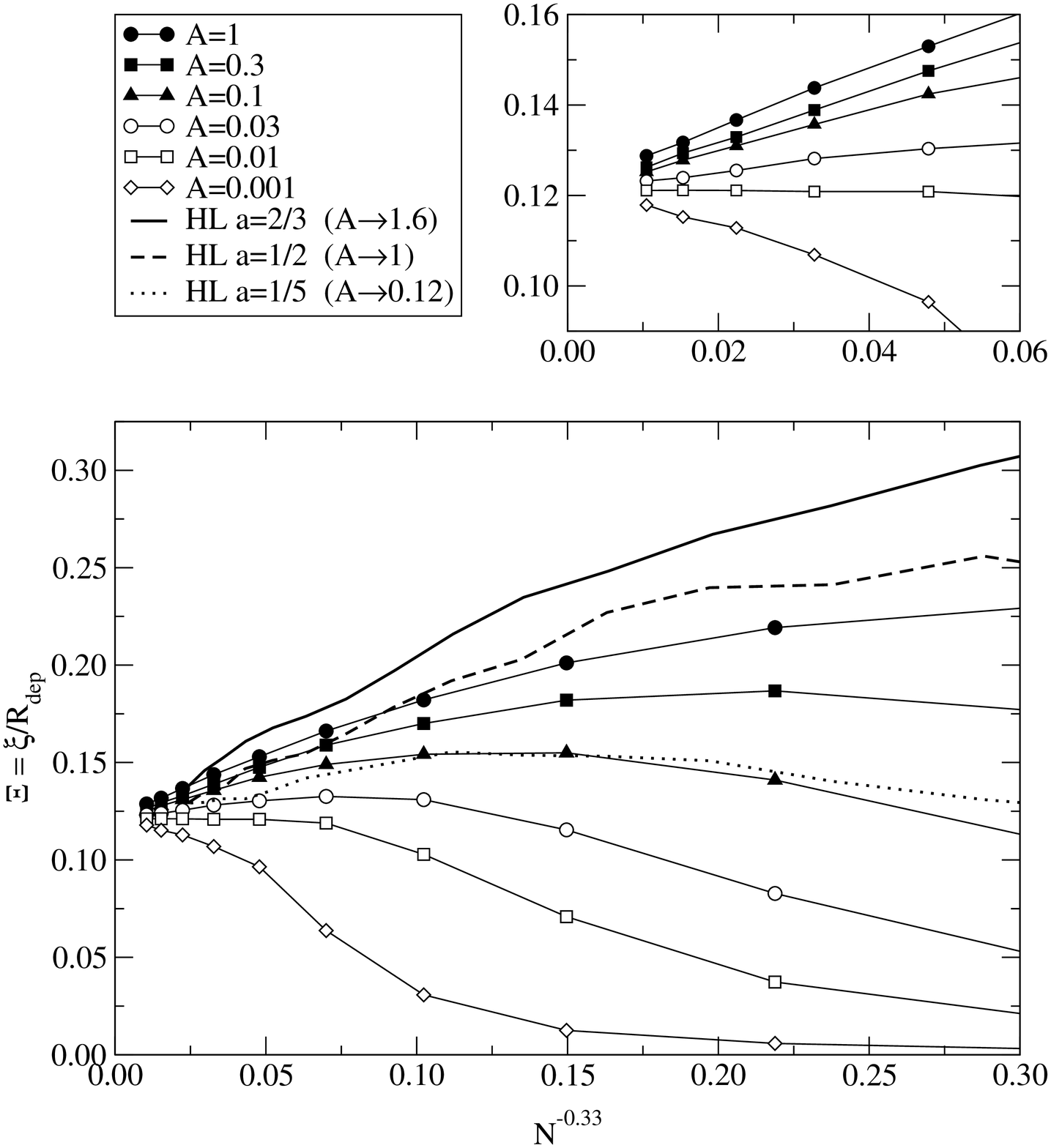}}
\medskip
\caption{Behaviour of the relative penetration depth $\Xi = \xi/R_{\mathrm
dep}$, with varying cluster size at various levels of noise reduction.  The
abscissa is chosen according to the correction to scaling exponent measured
from Fig.~\ref{differential xi}.  The relative penetration depth clearly
converges to a non-zero common value. Also shown are curves for clusters grown
by the Hastings-Levitov (HL) method with expected equivalence indicated in the
legend. The top right panel is a magnification of the asymptotic end of the
curves.}
\label{simple xi}
\end{figure}

Fig.~\ref{simple xi} shows a primary test of scaling:  how the
relative penetration depth, the ratio of penetration depth to mean radius of
deposition $\Xi \equiv \xi/R_{\mathrm dep}$, varies with $N$.  The different
levels of noise reduction are all consistent with a universal asymptote,
$\Xi_\infty\approx 0.12$, and with $N^{-0.33}$ as the common correction to
scaling at large $N$.
Fig.~\ref{simple xi} also shows data obtained with the HL iterated conformal
map method \cite{hastings}.  We can make a naive geometric argument to see how
the HL bumps correspond to different levels of off-lattice noise reduction. In
the HL method the bumps are generated by a conformal map parameterized by $a$
(see \cite{hastings} for details), with small $a$ giving shallow bumps.  Using
the scheme of Fig.~\ref{how dla works}, and working out when the aspect ratios
of the two types of bumps match, we find: $(1/a -1)^2 = 2/A -1$. Thus $a=0.5$
corresponds to ordinary DLA. Other equivalent
cases are indicated in the legend of Fig.~\ref{simple xi}, showing that this
naive argument represents reasonably well the relationship between our noise
reduction and that of Ref.~\cite{stepanov}. 

\begin{figure}
\epsfxsize=1\columnwidth
\centerline{\epsffile{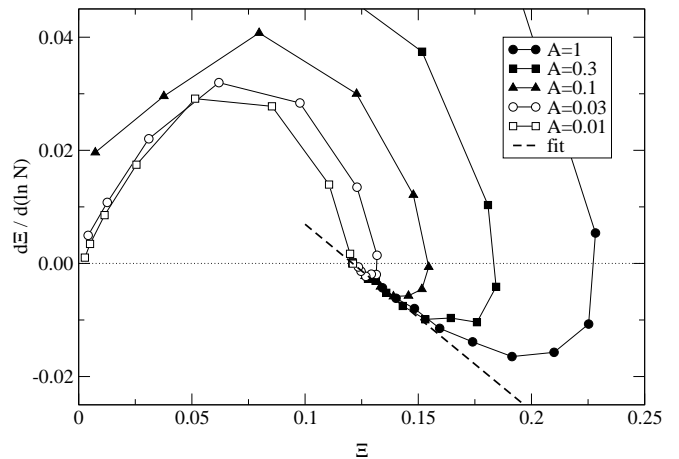}}
\medskip
\caption{Rate of change of the relative penetration depth $d \Xi/ d \ln(N)$
plotted against $\Xi$. The common dashed asymptote indicates that $\Xi$ has a
dominant correction to scaling of the form $\Xi = \Xi_\infty(1 + C N^{-\nu})$
with $\Xi_\infty = 0.121 \pm 0.003$ from the intercept of the plots and
$\nu=0.33 \pm 0.06$ from the slope.}
\label{differential xi}
\end{figure}

The correction to scaling exponents we report in this paper are not arbitrary 
fits, but directly measured as follows.  If we posit a leading asymptotic form
of some quantity, $Q$:
\begin{equation}
Q(N)=Q_\infty(1+C N^{-\nu})
\end{equation}
then a plot of $dQ(N)/d \ln(N)$ vs $Q(N)$ should have an intercept on the $Q$
axis of $Q_\infty$ approached with slope $-\nu$,  both independent of the
magnitude of $C$.  Fig.~\ref{differential xi} shows this analysis applied to
$\Xi$ and this is the basis for the choice of exponent $\nu=0.33$ for
Fig.~\ref{simple xi}.  This provides unbiased evidence that all the different
levels of noise reduction approach the same asymptotic value $\Xi_\infty$,
consistent with a common correction to scaling exponent.  Interestingly the
`fixed point' can be approached from either side,  corresponding to opposite
signs of $C$.

\begin{figure*}
\parbox{17.4cm}{
\parbox{1\columnwidth}{
\epsfxsize=1\columnwidth
\epsffile{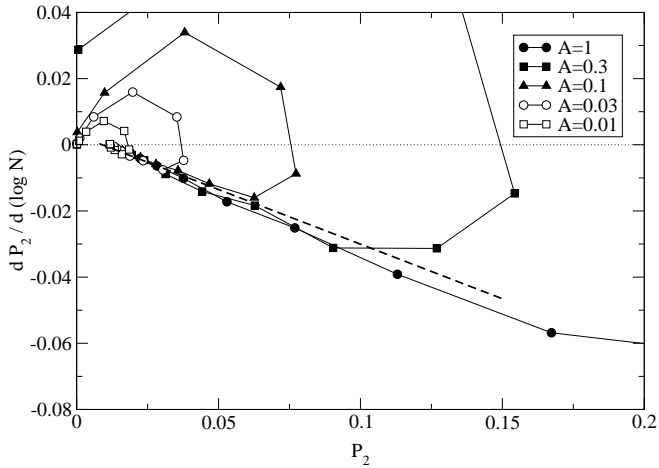}}
\parbox{1\columnwidth}{
\epsfxsize=1\columnwidth
\epsffile{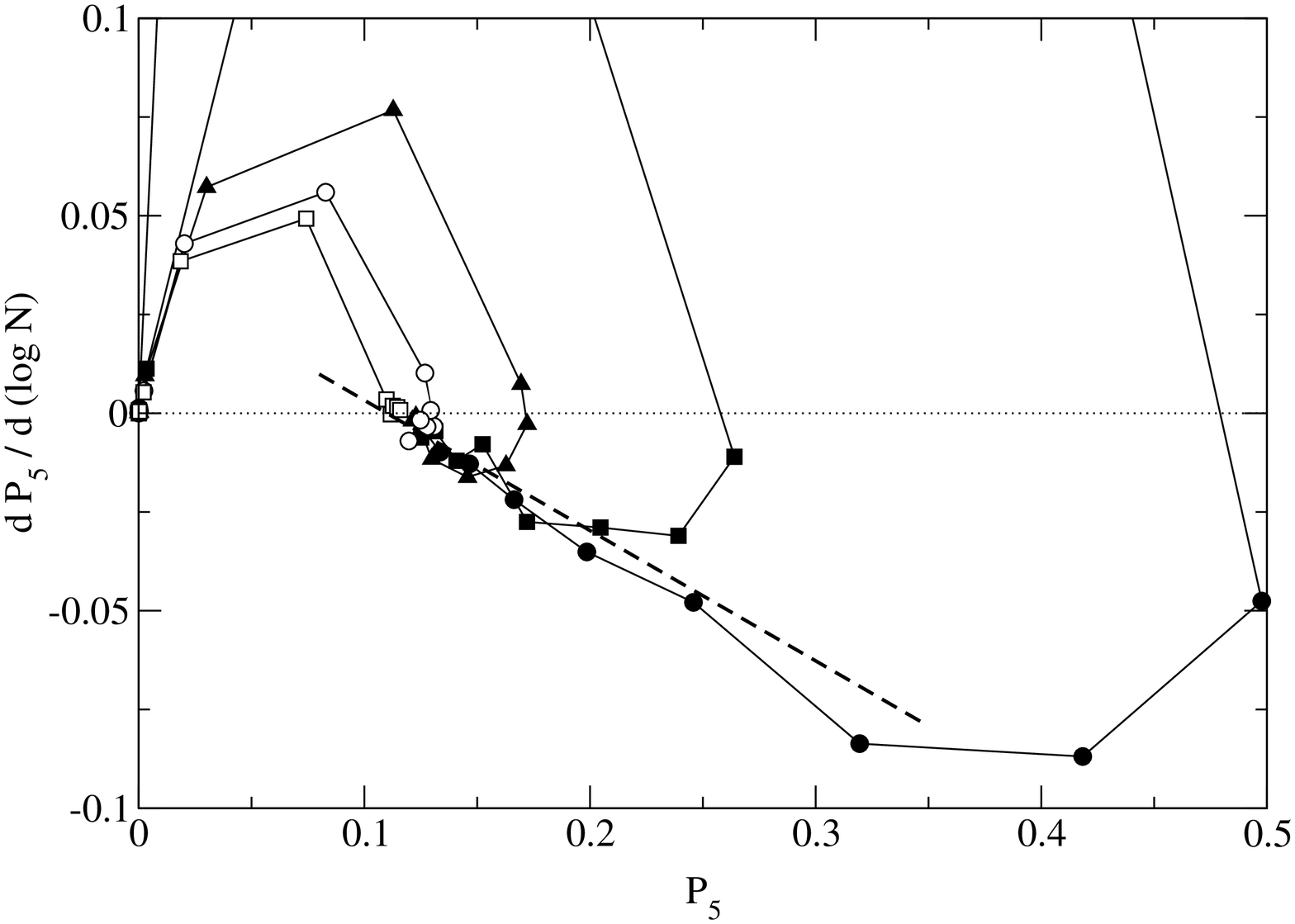}}
}
\medskip
\caption{The rate of change of the multipole powers $P_2$ and  $P_5$ with
respect to $\ln(N)$ against the multipole powers.  Each show a correction to
scaling exponent of 0.33 (dotted lines), within statistical error.}
\label{moments}
\end{figure*}

A deeper test of the universality of these clusters and their scaling comes
from the multipole moments of the growth probability distribution.  The $n$th 
multipole moment is given by
\begin{equation}
M_n = \int dq \, (x+iy)^n 
\end{equation}
where $q$ is the probability distribution for where growth will next occur.
(Note that $q$ is equivalent to the charge density on the  cluster surface
when it is considered to be a conductor held at a fixed potential). 
The multipole moments for positive $n$ fully  characterise the cluster shape,
and can be related invertibly to the Laurent coefficients of its conformal map
from the unit circle \cite{davidovitch}.  In practice we measured the $M_n$ by
sampling $(x+iy)^n $ with non-growing probe particles.

\begin{table}[b]
\caption{Best fit scaling exponents for $P_2$ to $P_5$. We have also measured 
$P_6$ to $P_{10}$. These yield somewhat larger apparent exponents with large
statistical errors.} 
\medskip
\begin{tabular}{ccccc} 
 \    & $P_2$            & $P_3$            & $P_4$            & $P_5$ \\\hline
$\nu$ &  $0.41 \pm 0.08$ &  $0.27 \pm 0.06$ &  $0.41 \pm 0.12$ &
 $0.40 \pm 0.12$ \\ 
\end{tabular}
\end{table}

\begin{figure}
\epsfxsize=1\columnwidth
\epsffile{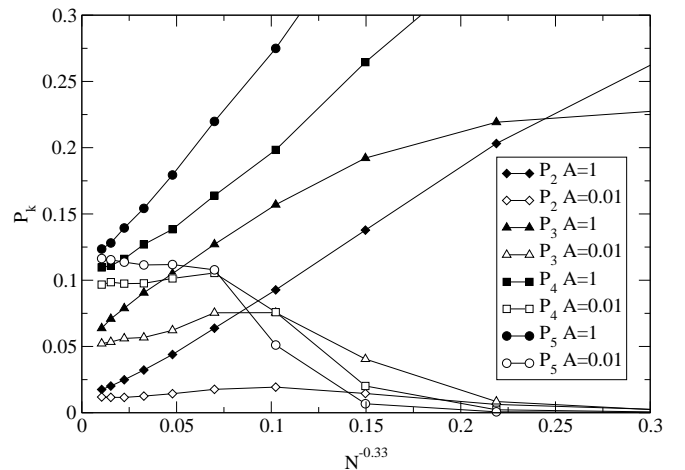}
\smallskip
\caption{Finite size scaling plots for $P_2$ to $P_5$ for $A=1$ and
$A=0.01$.  All the multipole powers exhibit the same correction
to scaling exponent.}
\label{Collected moments}
\end{figure}

Fig.~\ref{moments} shows the correction-to-scaling analysis of the
corresponding multipole powers 
\begin{equation}
P_n=\frac{|M_n|^2}{R_{\mathrm eff}^{2n}} \,,
\end{equation}
where we have scaled each $M_n$ by the appropriate power of the effective (or
Laplace \cite{davidovitch}) radius, $R_{\mathrm eff}$, which is given by 
$\ln R_{\mathrm eff} = \int dq \ln(r)$.  Each of $P_2$ to $P_5$ is
consistent with having a universal non-zero asymptote,  and moreover they are
all compatible with a single common correction to scaling exponent $0.33$, see
Table I.  Fig.~\ref{Collected moments} collects the resulting finite size
scaling plots assuming this exponent.  Together with the relative penetration
depth results, this presents strong evidence for universal asymptotic geometry
for DLA clusters, and a universal leading correction to scaling exponent $\nu
= 0.33$.

\begin{figure}
\epsfxsize=1\columnwidth
\epsffile{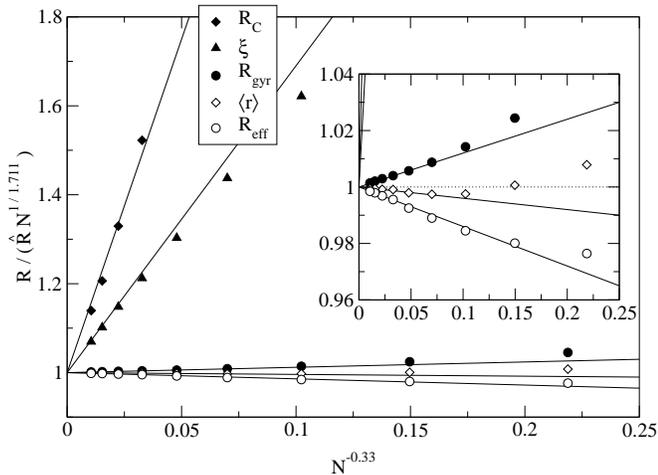}
\smallskip
\caption{Correction to scaling plots of various quantities of dimension length
(without noise reduction, $A=1$).  The largest correction is obtained by the
penetration depth $\xi$.  The inset magnifies the $y$ axis around 1.  }
\label{fig:rad-fit}
\end{figure}

In all of the measurements discussed above, the cluster center used was the
``center of charge'', natural to a snapshot of the growth. In the following
Section, however, we will require to compare data at different cluster sizes
where it becomes natural to use a fixed center, namely the cluster ``seed''.
Accordingly we have also measured the finite size scaling of various lengths
with the seed as fixed origin, and in all cases using direct ensemble
averages and for clusters with no noise reduction ($A=1$).
Using the seed as center also naturally leads to the measurement of
penetration depth as the rms spread of deposition radius about its ensemble
average:
\begin{equation}
\xi_0 = \sqrt{\langle r^2\rangle - \langle r\rangle^2}
\end{equation}
rather than computing the variance cluster-by-cluster before averaging, i.e.
\begin{equation}
\xi = \sqrt{\left\langle \int dq\,\,r^2 - \left(\int dq\,r\right)^2
  \right\rangle} \,\,.
\end{equation}
Fig.~\ref{fig:rad-fit} shows the fits of the following form:
\begin{equation}
R(N) = \hat R N^{1/D} (1 + \tilde R N^{-\nu}) \,,
\label{eq:rad-fit}
\end{equation}
and the coefficients are collected in Table~\ref{tab:rad-fit}.

It is worth noting that the effect of changing center is negligible except
perhaps for the penetration depth, where for $\xi$ using the center of of
charge (as for Figs.~\ref{simple xi} and \ref{differential xi}), the
coefficients are $\hat\xi=0.089$ and $\tilde\xi=5.8$ as opposed to those shown
for $\xi_0$ in Table~\ref{tab:rad-fit}.

\begin{table}
\caption{Coefficients of correction to scaling fits of form
Eq.~(\ref{eq:rad-fit}), with $D=1.711$ and $\nu=0.33$. The various lengths are
radius of deposition $R_{\mathrm dep}=\langle r\rangle$,
seed to center of charge distance $R_{\mathrm C} = \sqrt{\langle|\int dq\,
  {\mathbf r}|^2\rangle}$,
effective radius $R_{\mathrm eff}=\exp \langle\ln r\rangle$,
gyration radius $R_{\mathrm gyr} = \sqrt{\frac{1}{N}\sum_{N'=1}^N\langle r^2
  \rangle_{N'}}$, and
ensemble penetration depth $\xi_0=\sqrt{\langle r^2\rangle-\langle r\rangle^2}$,
where the averages are over the ensemble of clusters at fixed $N$.}
\medskip
\begin{tabular}{cddddd}
 & $R_{\mathrm dep}$ & $R_{\mathrm C}$ & $R_{\mathrm eff}$ & 
$R_{\mathrm gyr}$ & $\xi_0$ \\ \hline
$\hat R$   &  0.733 & 0.027 &  0.726 & 0.501 & 0.091 \\
$\tilde R$ & -0.04  & 15.  & -0.14  & 0.12  & 6.9
\end{tabular}
\label{tab:rad-fit}
\end{table}

\section{Multiscaling}
\label{sec:multiscaling}

Now we consider the anomalous scaling claim of multiscaling, when the
aggregate has a fractal dimension which depends upon distance from the seed as
a fraction of cluster radius. It was proposed in Ref.~\cite{coniglio} that the
ensemble average of the density of particles $g_N(r)$ of an $N$-particle
cluster at distance $r$ away from the seed obeys
\begin{equation}
g_{N}(xR_{\mathrm gyr}) = A(x) R_{\mathrm gyr}^{-d+D(x)} \,,
\end{equation}
where the dimension $D(x)$ is function of $x=r/R_{\mathrm gyr}$, and the size
$N$ and (average) radius of gyration $R_{\mathrm gyr}$ are of course
mutually dependent. Using the above formula at fixed $x$, one can extract the
dimension $D(x)$ by the scaling with $R_{\mathrm gyr}$:
\begin{eqnarray}
-d+D(x) &=& \left.\frac{\partial\ln g_{N(R_{\mathrm gyr})}(xR_{\mathrm gyr})}
  {\partial\ln R_{\mathrm gyr}}\right|_x \label{eq:Dx} \\ \nonumber
&=& \frac{R_{\mathrm gyr}}{dR_{\mathrm gyr}/dN}
  \left.\frac{\partial\ln g_N(xR_{\mathrm gyr}(N))}{\partial N}\right|_x
\end{eqnarray}
Simple fractal scaling would require $D(x)=D$ independent of $x$, but the
dimension measured this way in Ref.~\cite{amitrano} using medium size clusters
($N=10^4\ldots 10^5$) was observed to be a non-trivial function (see
Fig.~\ref{fig:multiscaling}).  Others partly confirmed that findings, although
with mixed results \cite{ossadnik91,ossadnik93}.

Now we will repeat the same measurement procedure but instead of
direct simulation we use the correction to scaling results of the previous
Section, within a scaling function assumption (see below). This turns out to
agree quantitatively with the earlier published $D(x)$ data, but implies that
the ultimate behaviour is simple fractal scaling with $D(x)\to D$ for all $x$.

\begin{figure}
\epsfxsize=1\columnwidth
\epsffile{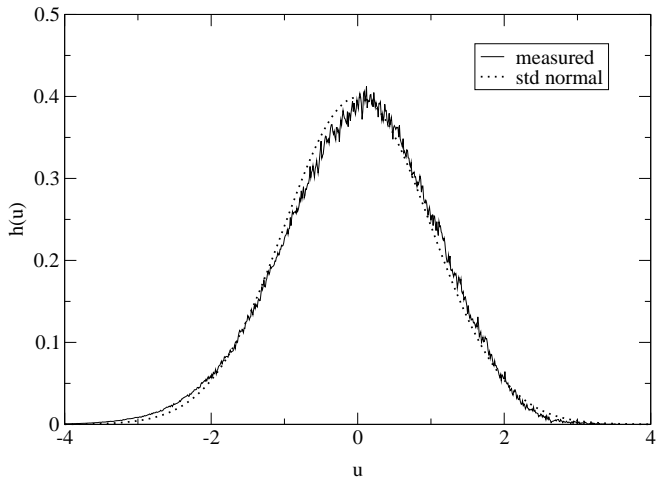}
\smallskip
\caption{The scaling function $h$. The measured data at $N=10^4$ (continuous
line) is compared to standard normal distribution (smooth dotted line). For
the measurement, the histogram bin width was $\Delta u = 0.01$.}
\label{fig:h}
\end{figure}

\begin{figure}
\epsfxsize=1\columnwidth
\epsffile{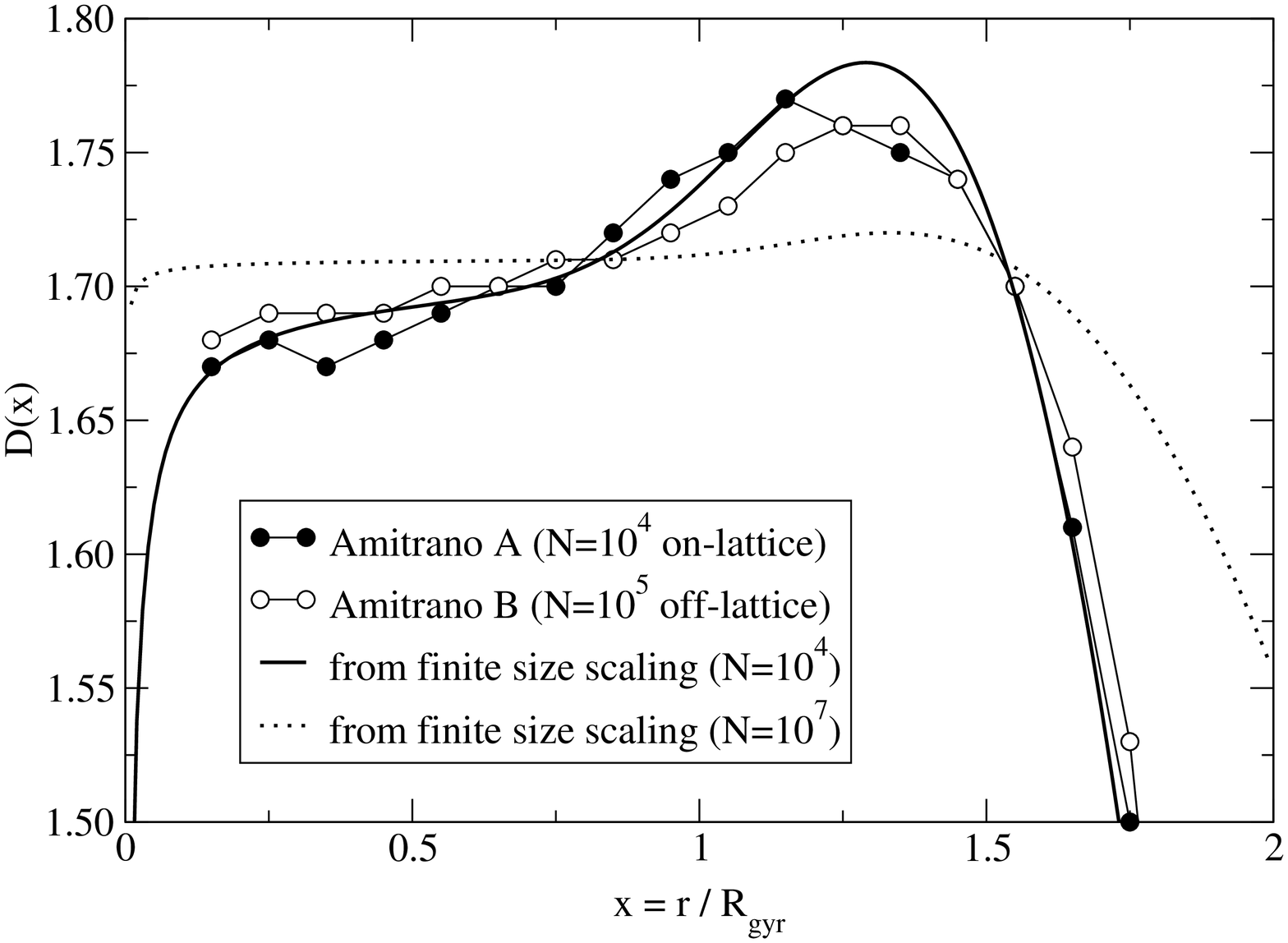}
\smallskip
\caption{Comparison of ``multiscaling dimensions'' from
Ref.~\protect\cite{amitrano} and the finite size scaling prediction discussed
in the text. The finite size scaling prediction implies that as $N\to\infty$,
$D(x)\to D$ for all $x$, and the predicted approach at $N=10^7$ is shown. The
only inputs to the finite size scaling curves are $R_{\mathrm dep}$, $R_{\mathrm
gyr}$ and $\xi_0$ using Eq.~(\ref{eq:rad-fit}) with parameters from
Table~\ref{tab:rad-fit}, plus a Gaussian model for the scaling function (see
Fig.~\ref{fig:h}).}
\label{fig:multiscaling}
\end{figure}

Consider the distribution of $r$, the distance of attaching particles from the
seed: as we have seen, this has mean $R_{\mathrm dep}$ and variance $\xi_0$.
Now we assume that the {\em shape} of the probability density function is
independent of $N$:
\begin{equation}
\frac{1}{\xi_0(N)}\,\,h\!\left(\frac{r-R_{\mathrm dep}(N)}{\xi_0(N)}\right) \,,
\end{equation}
where $h$ is a normalized probability density with zero mean and unit
variance. After replacing the sum over particles with an integral, for the
particle density we get
\begin{equation}
g_N(r) = \int_0^N \frac{dN'}{\xi_0(N')}\,\,
  h\!\left(\frac{r-R_{\mathrm dep}(N')}{\xi_0(N')}\right) \,.
\label{eq:ouransatz}
\end{equation}
A similar formula has been suggested in \cite{lee}. Given that we have already
studied $R_{\mathrm dep}(N)$, $R_{\mathrm gyr}(N)$ and $\xi_0(N)$, the only
outstanding quantity to be found is the scaling function $h$, which we find
to be very close to the standard normal distribution, see Fig.~\ref{fig:h}.

Fig.~\ref{fig:multiscaling} shows how well $D(x)$ derived from our finite size
scaling results plus a normal distribution for $h$ agrees with the raw data of
Ref.~\cite{amitrano}. Also shown is what our results imply for the behaviour
at larger $N$, and as $N\to\infty$ with $R_{\mathrm dep}$, $R_{\mathrm
gyr}$ and $\xi_0$ approaching pure scaling, $D(x)\to D$. Thus we conclude that
all the apparent reported $x$-dependence of $D(x)$ arises from corrections to
scaling, and indeed almost all the effect cames from the relatively large
corrections to scaling in $\xi_0$. Our new interpertation of this data also
resolves a previously noted paradox \cite{amitrano}, namely that $D(x)$
increasing with $x$ cannot be asymptotic scaling as it would imply some
decrease of $g_N(r)$ with increasing $N$ at fixed $r$.

\section{Size fluctuations and fixed point}

We now present an interpretation of the leading correction to scaling, based
on new observations from our data and building on earlier work \cite{barker}.
The amplitude of the leading correction to scaling crosses zero at a common
value of noise reduction $A_f\approx 0.01$, for {\em all} of the plots in
Figs.~\ref{simple xi} and \ref{Collected moments}. This suggests that the
noise reduction and the correction to scaling are fundamentally related, which
can be understood by using the renormalization theory of noise reduction of
Barker and Ball\cite{barker}. In this view, the cluster is approximated as
being at its large $N$ fixed point if one unit of growth acts as a coarse
graining of DLA on finer length scales. This seems to occur if we grow with
input noise near $A_f$. This is equivalent to have $\delta N/N=\sqrt{A_f}$ for
relative fluctuation in the number of particles to advance the growth locally
by one particle diameter.

\begin{figure}[!b]
\epsfxsize=1\columnwidth
\epsffile{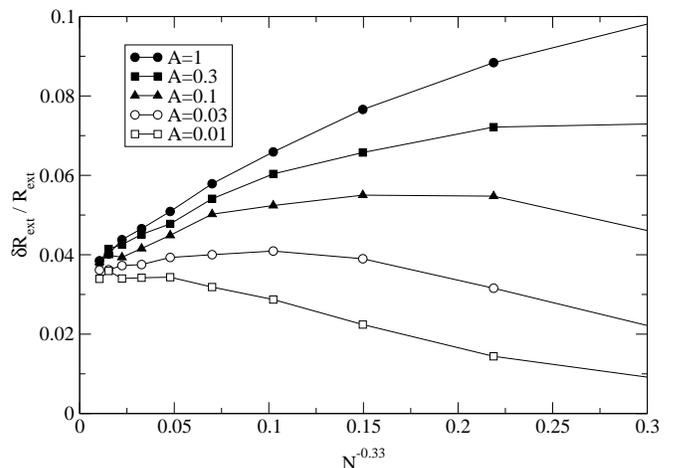}
\smallskip
\caption{Ensemble spread of extremal cluster radius, which tends towards the
value $0.035 \pm 0.003$.}
\label{Rext}
\end{figure}

\begin{figure}
\epsfxsize=1\columnwidth
\epsffile{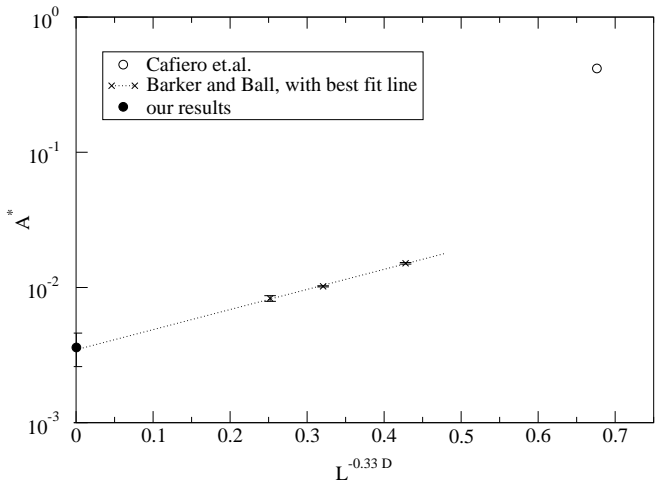}
\smallskip
\caption{Estimate of the fixed point value of $A$ ($A^*$), combined with
previous estimates from Barker and Ball (the middle three points) and Cafiero
(the rightmost point).  The Barker and Ball data is in good agreement with our
results, but the Cafiero data disagrees.}
\label{A Fixed}
\end{figure}

We can also view this in terms of a fixed point for the noise output of the
growth, $\sqrt{A_{\mathrm out}}=\delta N/N$, in terms of the relative
fluctuation in the number of particles to span a fixed radius.
Fig.~\ref{Rext} shows our data for the ensemble
spread of extremal cluster radius.  Since this spread is small, we can infer:
\begin{equation}
\left. \frac{\delta N}{N} \right| _{R_{\mathrm ext}} = D \left. \frac{\delta 
R_{\mathrm ext}}{R_{\mathrm ext}}\right| _{N}
= 0.060 \pm 0.005
\end{equation}
from our extrapolated value.  Thus we find an asymptotic renormalised noise
$A^* = 0.0036 \pm 0.0006$. This is in qualitative agreement with our observed
value of $A_f$. Furthermore, Fig.~\ref{A Fixed} shows how well this
vindicates Barker and Ball's earlier estimates of the fixed point,  using our
value of $\nu$ to extrapolate from their finite size calculations.  By
contrast, the more recent work of Cafiero {\it et al}\cite{cafiero} using a
very small scale renormalisation scheme disagrees by two orders of magnitude.

Our interpretation is thus that the renormalized noise is the slow variable
which dominates convergence of other quantities to scaling.   Our observed
input noise value of $A_f \simeq 10^{-2}$ (for the leading correction to
scaling to vanish) and the extrapolated fixed point output noise $A^*$ are
equal within a factor of order unity, showing the consistency of the picture.

We can take this
interpretation a step further to infer that the dominant fluctuations of 
$R_{\mathrm ext}$ determining
the noise reduction are fluctuations in cluster shape rather than 
overall cluster radius.  The basis for this is that the
logarithmic average radius, $R_{\mathrm eff}$,  has much smaller 
spread, asymptotically $\delta R_{\mathrm eff}/R_{\mathrm eff}= 0.012 \pm 0.001$
compared to $\delta R_{\mathrm ext}/R_{\mathrm ext}=0.035 \pm 0.003$.  
Since $R_{\mathrm eff}$
is an average which emphasizes typical size,
the larger fluctuations in $R_{\mathrm ext}$ which gave us
$A^*$ must be attributed to shape.  (However, we showed in
\cite{somfai} that $R_{\mathrm eff}$ has the same crossover exponent, $\nu$, as the other
quantities discussed here.)
In this sense DLA clusters are fundamentally
stochastic objects with a distribution of shape. 

\section{Summary}

We believe our work opens the way to a definitive view of DLA in two dimensions,
and the extension of this work to three dimensions is in hand.  The 
identification of `DLA fixed point behaviour' is now reasonable,  as we have 
shown the sort of universal limiting amplitudes and correction to scaling 
exponents associated with such terminology.  

Some main areas are outstanding. First, the renormalized noise, $\delta N/N 
= \sqrt{A^*}$, is not
of order unity, as we might expect \emph{a priori}, and as has been
suggested \cite{cafiero}. We do not understand
the origin of this small number, and tracing its origin is a 
central remaining challenge in understanding DLA. 
Another puzzle which we hope to address in a later paper is why the fractal
dimension is comparatively insensitive to the convergence of the renormalized
noise.

Also, we need to understand the full scaling of the probability distribution
for growth in DLA, corresponding to the harmonic measure of the perimeter.  To
this end the more expensive cluster growth methods of Hastings and
Levitov\cite{hastings} (HL) are likely to come into their own as they yield the
harmonic measure directly.  Stepanov and Levitov\cite{stepanov} have already
shown some results for HL clusters grown with shallow bumps, corresponding
rather closely to our noise reduction technique.  

The richer, simpler area to explore is the response to anisotropy and its
sensitivity to noise.  Small DLA clusters appear robust to the intrinsic bias of
growing on a square lattice, whereas large clusters (and equivalently noise
reduced ones) are driven to grow a four fingered dendrite.  The first
requirement is a systematic analysis of how this is a relevant perturbation of
the isotropic DLA fixed point.  Secondly, we might ask whether the anomalous
response for small simple DLA clusters is dominated by some other hitherto
unsuspected fixed point with much larger noise level.  There is another 
rather neglected nearby fixed point, that of spherical growth, which becomes 
more pertinent at high noise reduction - where it takes longer to exhibit 
its instability.  We suggest the influence of this fixed point may 
be responsible for shifting the observed $A_f$ somewhat above $A^*$,  and this 
should be relatively amenable to analytic theory. 

\acknowledgements

NEB would like to thank BP Amoco and EPSRC for the support of a CASE award
during this research. ES is supported by the Dutch FOM Foundation and the EU
Marie Curie Fellowship.  We thank Paul Meakin and Thomas Rage for sending us a
computer code which we used for some of the results presented here.

\bibliographystyle{unsrt}
\bibliography{References}

\end{document}